# ADAPTIVE FILTER DESIGN FOR STOCK MARKET PREDICTION USING A CORRELATION-BASED CRITERION

**J. E. Wesen**
State University of Amazon, Electrical Engineering Department,
Av. Darcy Vargas, 1200, Manaus – Amazon, 69065-020 Brazil
e-mail: ednelson.wesen@gmail.com

**V. Vermehren V.**
State University of Amazon, Electrical Engineering Department,
Av. Darcy Vargas, 1200, Manaus – Amazon, 69065-020 Brazil
e-mail: vvv@netium.com.br

**H. M. de Oliveira**
Federal University of Pernambuco
Rua Ac. Hélio Ramos, s/n 4º andar, 7800, Recife – Pernambuco, 50.711-970 Brazil
e-mail: hmo@ufpe.br  www2.ee.ufpe.br/codec/deOliveira.html

## RESUMO

Este trabalho apresenta uma nova abordagem usando filtragem adaptativa para a predição de ativos em bolsas de valores. Os conceitos introduzidos permitem a compreensão deste método e o cálculo da previsão correspondente. Este método é aplicado, em um estudo de caso, através da previsão da valorização real das ações PETR3 (Petrobrás ON), negociadas no mercado acionário brasileiro. As escolhas de primeira linha para o comprimento da janela e para o número de coeficientes de filtro são avaliadas. Isto é feito observando-se a correlação entre o sinal preditor e o curso real das ações no mercado, em termos de comprimento da janela de predição e os valores dos coeficientes do filtro adaptativo. Mostra-se que tais preditores adaptativos fornecem, em média, um lucro substancial sobre o montante investido.

## ABSTRACT

This paper presents a novel adaptive-filter approach for predicting assets on the stock markets. Concepts are introduced here, which allow understanding this method and computing of the corresponding forecast. This approach is applied, as an example, through the prediction over the actual valuation of the PETR3 shares (Petrobrás ON) traded in the Brazilian Stock Market. The first-rate choices of the window length and the number of filter coefficient are evaluated. This is done by observing the correlation between the predictor signal and the actual course performed by the market in terms of both the window prevision length and filter coefficient values. It is shown that such adaptive predictors furnish, on the average, very substantial profit on the invested amount.

## 1. Introduction

Adaptive filter is a device able to perform self-learning, that is, when the time goes, this filter set the output in conformity of the required performance. The main feature of this filter is the fact that they have skill to modify its response in real time with the intent to improve its performance. In practice, the adaptation algorithm is implemented through two classical methods, gradient method and least square (LMS algorithm, RLS algorithm). These filters have many applications ranging from filtering, spectral analysis, signal detection and equalization. A wide application area of this tool is in the future sequences prevision through using past sequences information (Prediction). Adaptive predictor has been used in Linear Predictive Code (LPC), Adaptive Differential Pulse Code Modulation (ADPCM) [Haykin (2009)].

Figure 1 shows a universal scheme for applying an adaptive filter in the prediction framework, where $k$ is the iteration number, $x(k)$ is input signal, $y(k)$ is the adaptive filter output what is an estimate about wanted response $d(k)$, and $e(k)$ is the error signal defined as the difference between the wanted response and the filter output, i.e., $e(k) := d(k) - y(k)$.

RLS (Recursive Least-Squares) adaptive algorithm offers high performance and high speed of convergence when running in time variant environments.

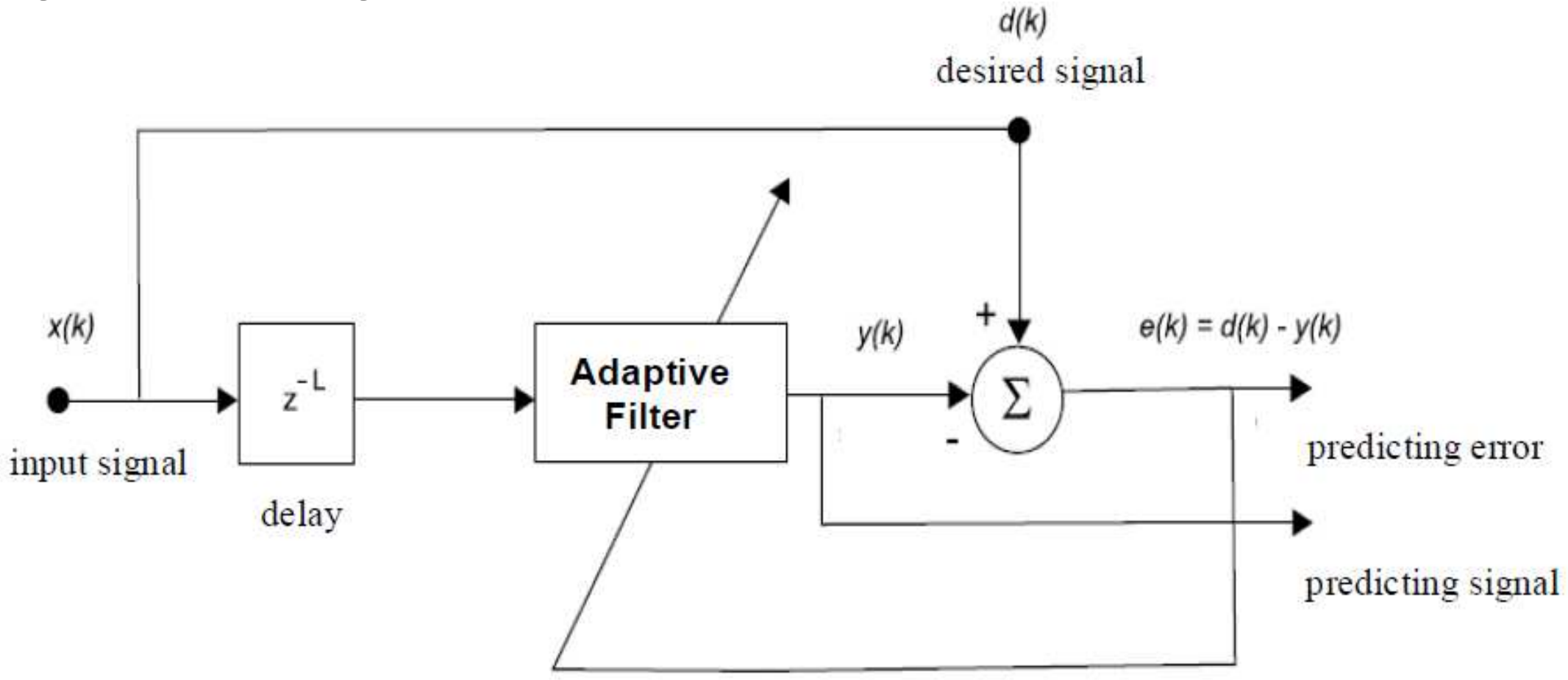

Figure 1. Adaptive filter general block diagram for signal prediction.

A window observation is the time period under analysis, of which usually has a rectangular or exponential fit. The parameter that controls the fit and duration of the observation window is called forgetting factor $\lambda$ (because it accounts for the memory of the algorithm), $0<<\lambda<1$ and the objective function is given by:

$$F(e(k)) = \xi^{d}(k) = \sum_{i=0}^{k} \lambda^{k-i} e^{2}(k) = \sum_{i=0}^{k} \lambda^{k-i} \left[ d(i) - x^{T}(i) w(k) \right]^{2}.$$

This function is convex in the multidimensional space $w(k)$, i.e., $\xi^{d}(k)$ has only one global minimum and no one local minimum. Therefore, we can achieve that point by setting the $\xi^{d}(k)$ gradient equals to zero, which results

$$w(k) = \left[ \sum_{i=0}^{k} \lambda^{k-1} x(i) x^{T}(i) \right]^{-1} \sum_{i=0}^{k} \lambda^{k-1} x(i) d(i).$$

There is a solution with computational complexity of order $O[N^2]$ for this problem [Diniz (2002)], but it may occur some algorithm instability.

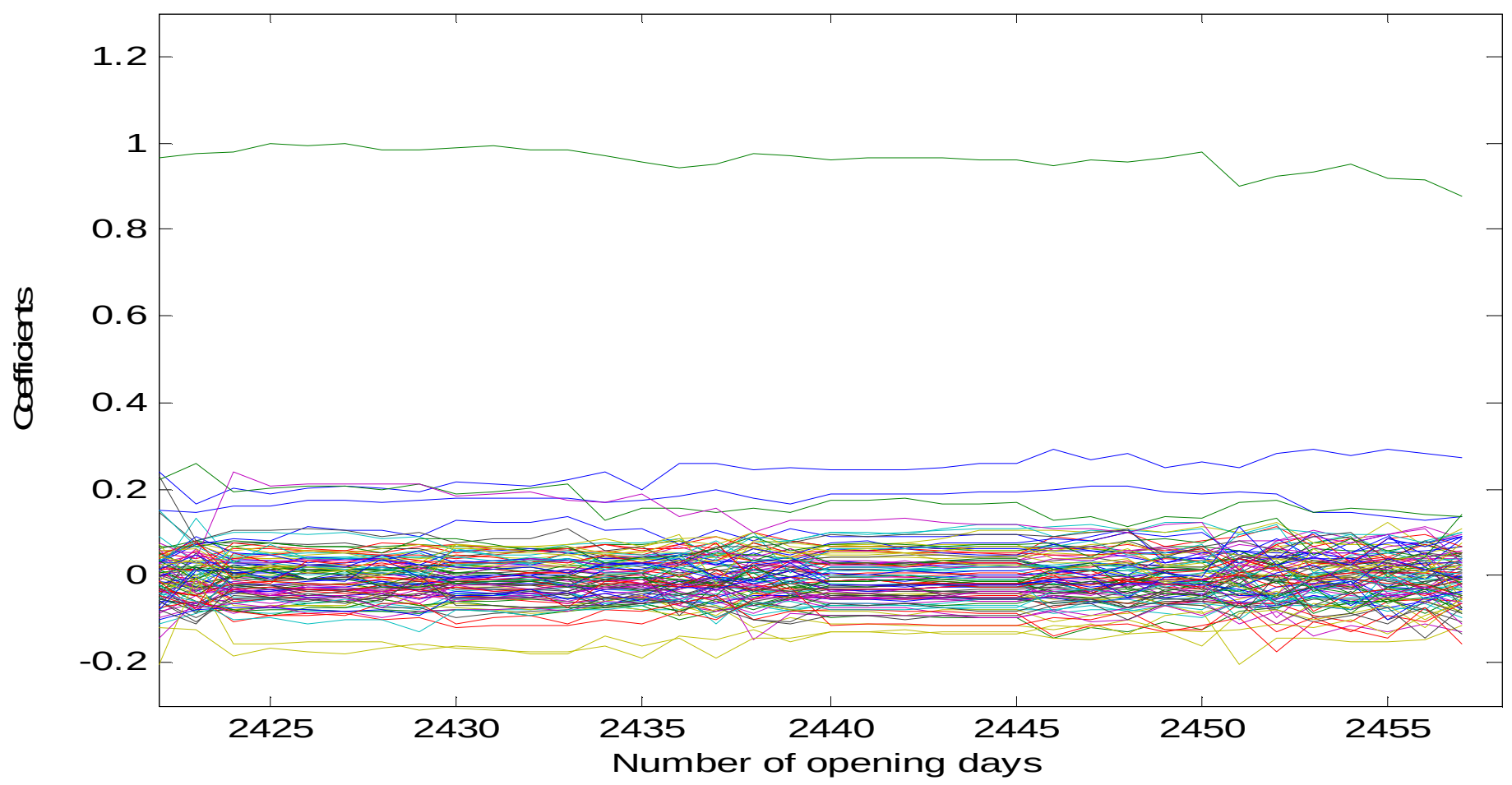

Figure 2. Convergence of the coefficients of the adaptive filter running to perform the prediction of the PETR3 (Brazilian stock market).

This work addresses the application of adaptive filters in an important environment, namely, stock market prediction [Cowles (1993), Antoniou (2004)]. The accuracy of prediction markets was investigated in [Berg (2003), Dokko (1989)]. Several tools have been suggested in the analysis of the tendency of shares on stock exchanges. The stock market forecasting can be carried out using ARIMA [Wang (1996), Pai (2005)] and further time series approaches [LeBaron (1999)], neural networks [Lawrence (1997)], hidden Markov chains [Turner (1989), Hassan (2007)], genetic algorithms [Werner (2002), Chen (2001)] and fuzzy sets-based models [Östermark (1996)]. Stochastic process modeling can also be useful for market share forecasting [Mandelbrot (1963)]. Even nonlinear models can be applied in the area [Franses (1996)]. Models using auto regressive and moving average (AR and ARMA models) are extrapolated from linear difference equations can be applied in the area too [Wan (1994)].

## 2. Application Adaptive Filter in the Framework of Stock Market Prediction

We intend to investigate the viability of the adaptive filtering as a tool for forecasting on the stock exchange. Application of LMS adaptive filter has already been successfully proposed for traffic forecasting over wireless networks [Liang (2002)]. Adaptive filtering was used to obtain an estimate of the price quotation of shares of Petrobrás traded on the Brazilian stock exchange Bovespa (*Bolsa de Valores de São Paulo*). For purposes of this study was chosen active Bovespa called PETR3 (Petrobrás ON) comprising a main driver of Brazilian stock market by more than 15% of Bovespa index [WWW (2011)].

The type of filter used in implementing the predictor for the active PETR3 was an adaptive FIR digital filter with 100 real coefficients. The adaptation algorithm is the RLS with forgetting factor of 0.98. A window of predictability of length 16 days was adopted. The simulation was performed in MATLAB™ platform [Kamen (1997)] corresponding to the period from 01/03/2000 until 09/23/2009. The values of daily prices correspond to the instant of closing the stock exchange. Fig. 2 shows the convergence of the coefficients of the adaptive filter running to perform the prediction of the PETR3.

The evolution of the filter coefficients shown in this figure clearly indicates that there is convergence in the period around the time of prediction. The abscissa axis here is the number of days of opening the stock exchange after the first day considered in the simulation, i.e. 01/03/2000. The actual valuation of the share PETR3 (Petrobrás ON) traded in the Brazilian Stock Market through adaptive filtering is shown in the solid line of Fig. 3 [WWW (2011)].

The forecast derived from the adaptive filter is also shown in the same figure.

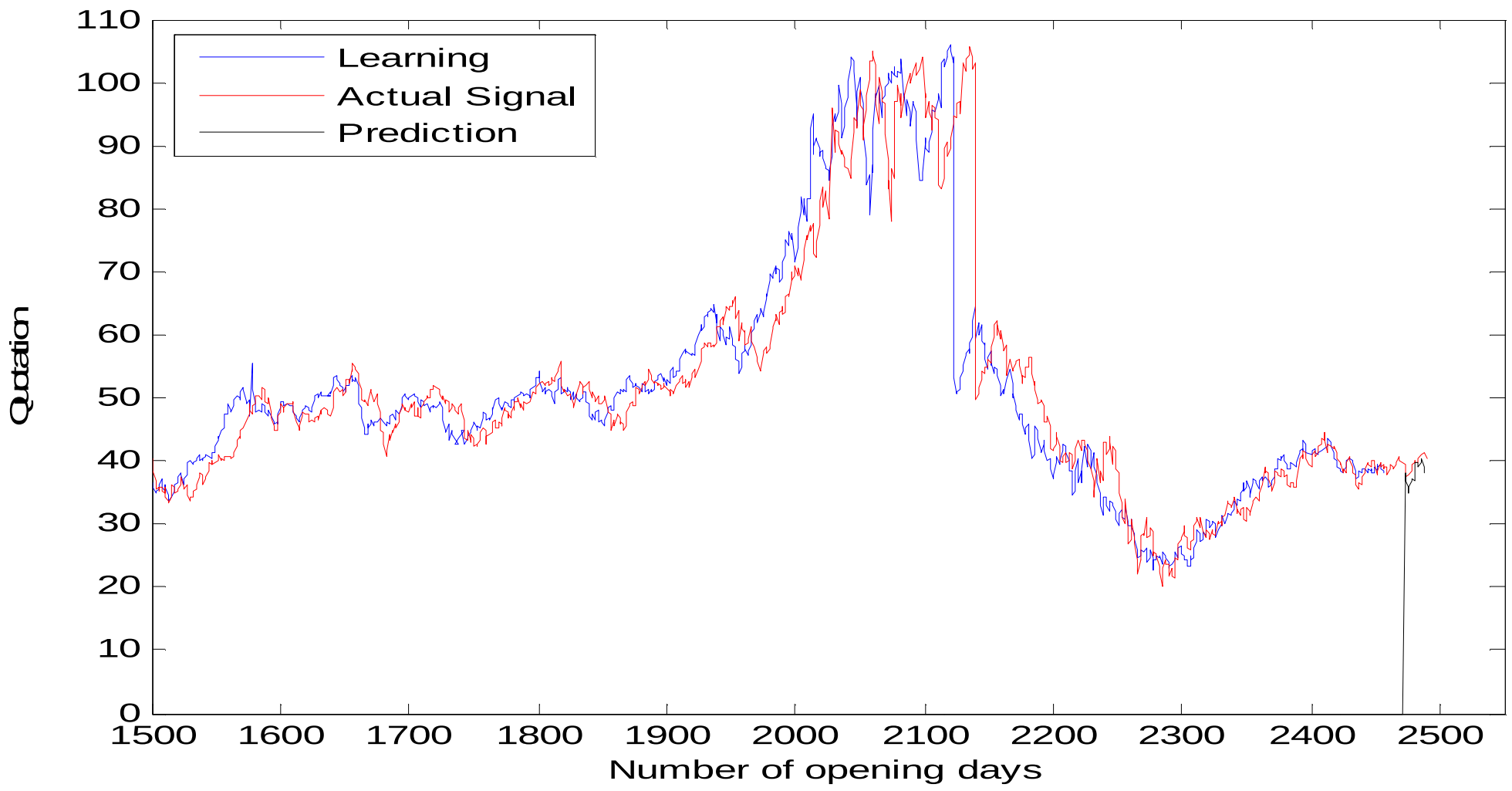

Figure 3. Valuation of the asset PETR3 (Petrobrás ON) in the Brazilian Stock Market through adaptive filtering: a) actual market trend b) guesstimate obtained from adaptive filtering.

It can be observed that the prediction from the adaptive filtering is in agreement with the actual price quotation evolution of Bovespa. In Fig.3-4, the prediction delay of 16 days, which corresponds to the prediction window, is clearly observed.

As an example, we performed the prediction for a specific day, namely, 2,459 overture days after 1/3/2000. Fig. 4 shows a zoom of the prediction made by the adaptive filter in this period, exhibiting further details about the behavior of the predictor.

## 3. Performance of the Adaptive Filter for Predicting a Representative Asset of the Bovespa

Let us consider the strategies for investment in the shares PETR3 using prediction provided by the adaptive filter from the moment 2473 (shown in Fig. 5)· The filtering is performed with the RLS algorithm using 100 coefficients and a forgetting factor $\lambda$ = 0.98. The estimate trend of the price quotation of PETR3 based filtering for a 16 days prediction window is displayed in dotted line of Fig 5.

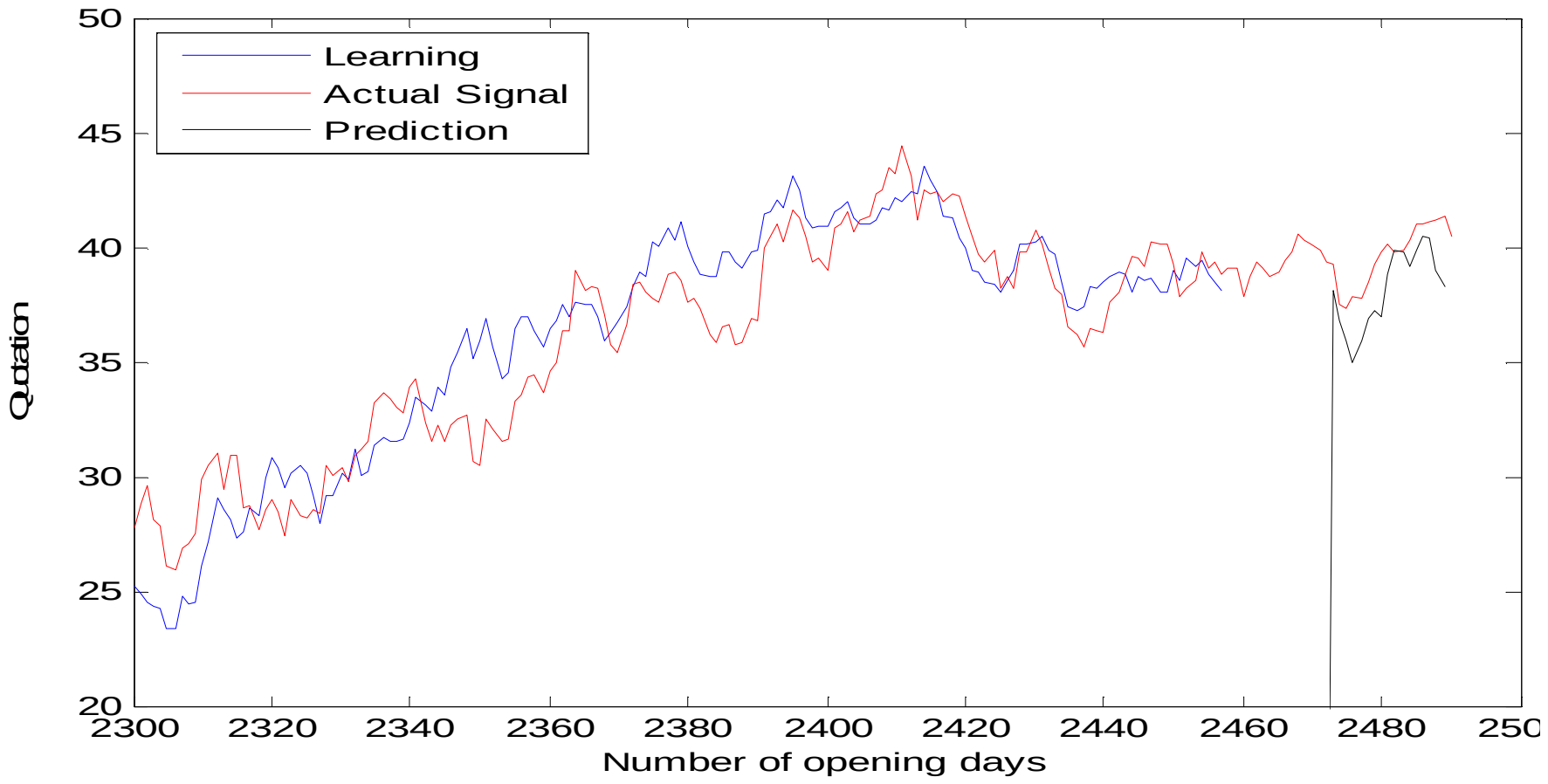

Figure 4. Behavior of predictions (Detail of Fig.2) around a particular day (2,459 overture day after 1/3/2000) where the estimative is performed.

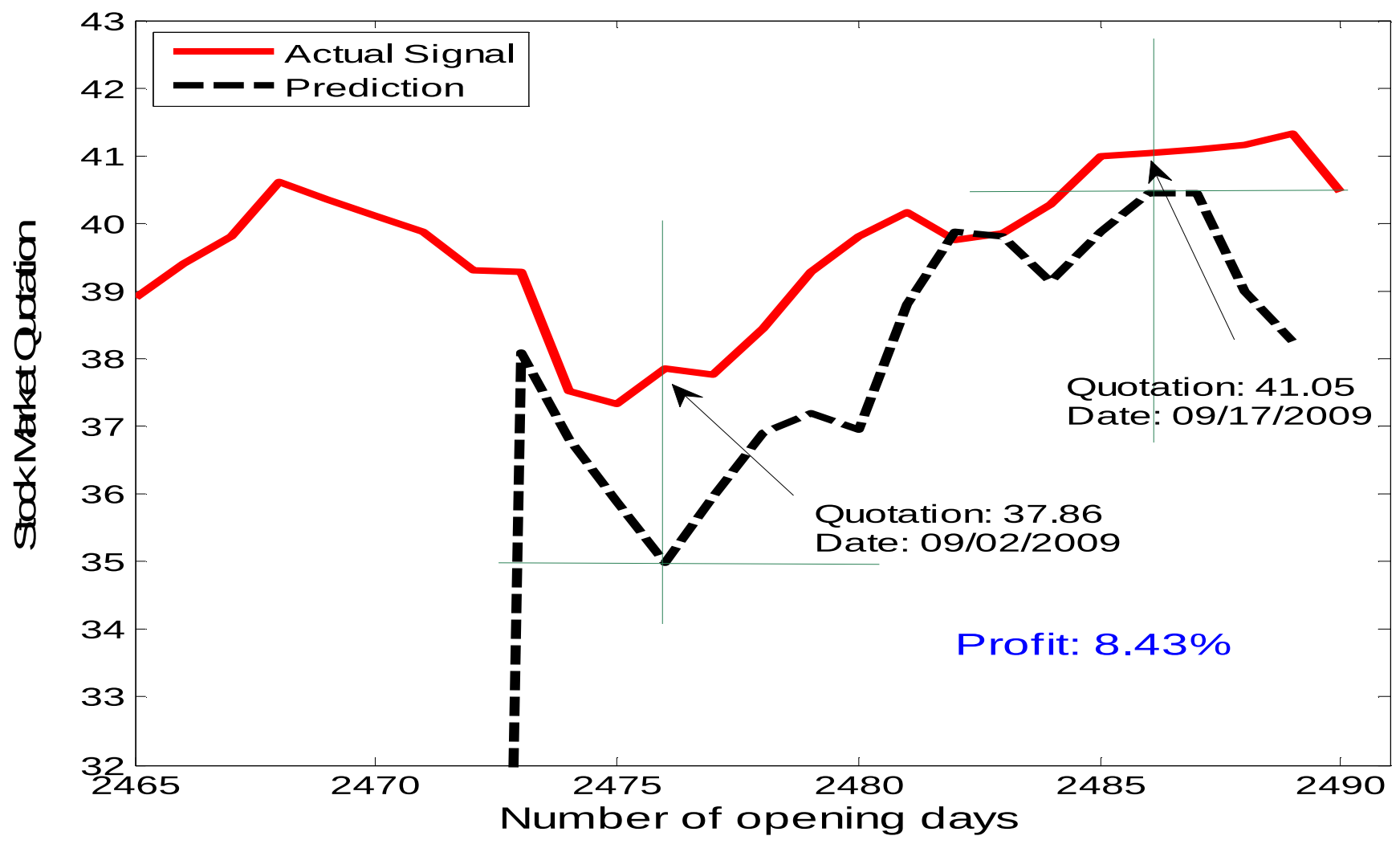

Figure 5. Market behavior of the active PETR3 between 08/18/2009 (2,465) and 09/22/2009 (2,489). a) Prediction according to adaptive filtering (dotted), b) Actual market quotation (solid).

The prediction curve suggests the acquisition of the asset on 09/02/2009 (buying in equities) and sell these shares at the moment 17/09/2009 (Sale of shares on the rise). The actual behavior indicated the solid line in Fig. 5 shows that the purchase would be accomplished with quote conducted with 37.86 and the sale price 41.05 [WWW (2011)]. If this strategy were adopted, this would allow a profit of 8.43% on the invested amount. Although this instance is quite particular, it illustrates the power of the adaptive-filter based prediction. Further days randomly chosen were also investigated, and the general results have been similar.

The application presented in this section illustrates the potential of this engineering technique in the decision making on investment strategies to be adopted in the stock market. The next section evaluates the influence of a few parameters of the adaptive filter in forecasting the quotation of PETR3. The robustness of the estimate produced by the adaptive predictor is also evaluated.

## 4. Adaptive Filter Design and Accuracy Analysis of Predictions

A number of challenges can be devised in the above mentioned scenario. To begin with, it could be asked for which type of assets this method performs better. What about the number of coefficients of the FIR should be used to improve the performance of the predictor? With respect to the prediction window, is the quality of prediction size-dependent? Has the value of the forgetting factor some influence about the results of the prediction? Some of these inquiries are answered in the sequel. In order to assess the performance of the adaptive filter estimation, the correlation was used as a similarity measure to investigate the goodness of fit between the predicted signal and the actual market quotation, within a given prediction window.

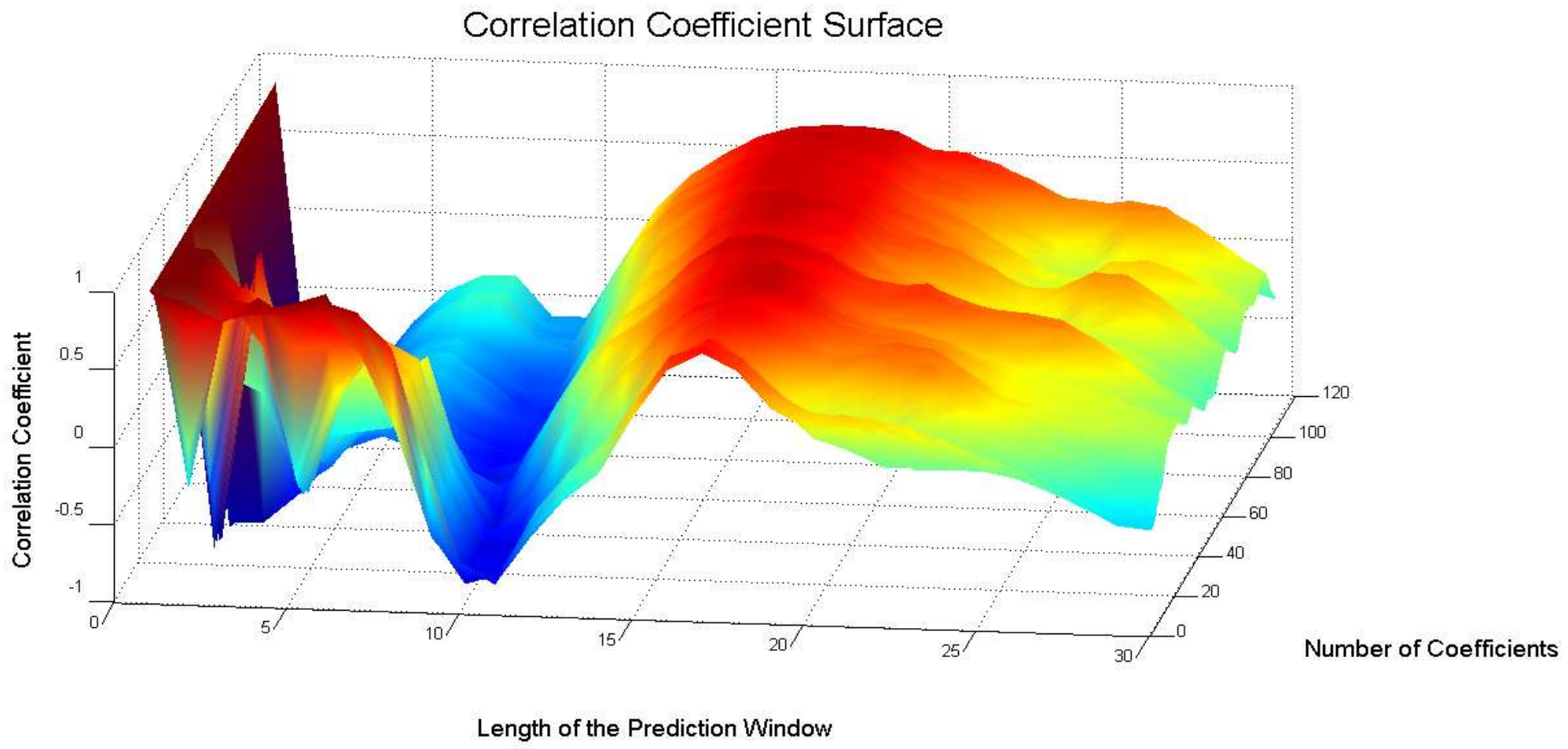

Figure 6. Correlation 3D-plot between the predicted signals obtained from the adaptive filter predictor and the actual market quotation.

The accuracy analysis was then carrying out by investigating the correlation values as a function of the number of coefficients of the filter (filter design) and the size of the prediction window. Figure 6 shows the 3D-correlation plot between 08/18/2009 (2,465) and 09/22/2009 (2,489).

Figure 6 was converted into a three-dimensional contour chart, which is presented in Fig. 7, where it can be realized the areas of an increased correlation between the predictor signal and the actual sequence performed by the market.

In order to investigate the influence of the number of coefficients of the FIR filter used in the adaptive predictor, the side profile of three-dimensional surface shown in Fig. 6 was drawn (Fig. 8), indicating the correlation peaks seen by an observer located on the right of the surface. It can be observed that the region between 30 and 100 coefficients presents a high correlation. We clearly distinguish that less than 30 FIR coefficients are not enough to get a fine prediction.

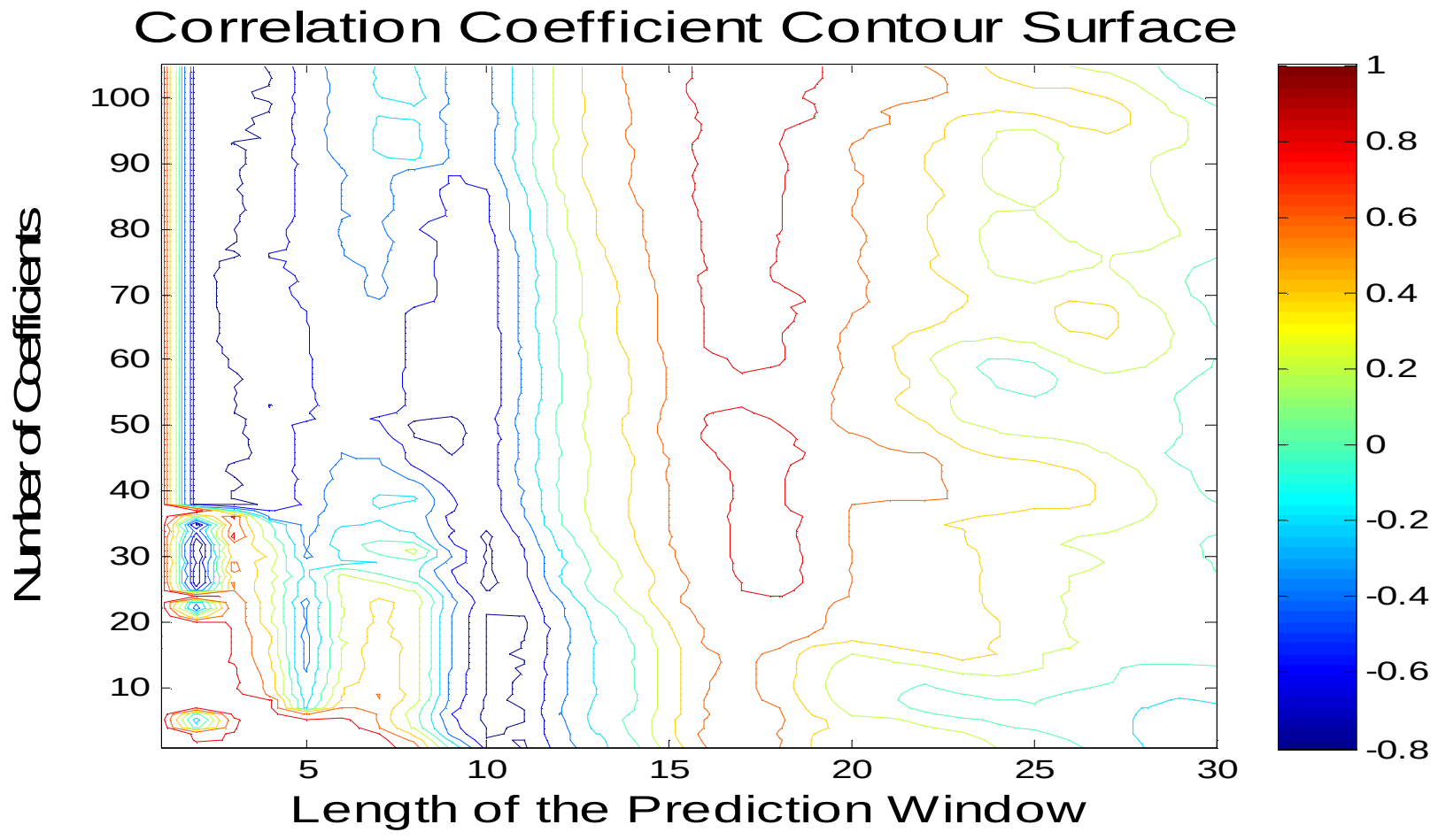

Figure 7. Contour plot corresponding to the correlation surface (see Fig.6).

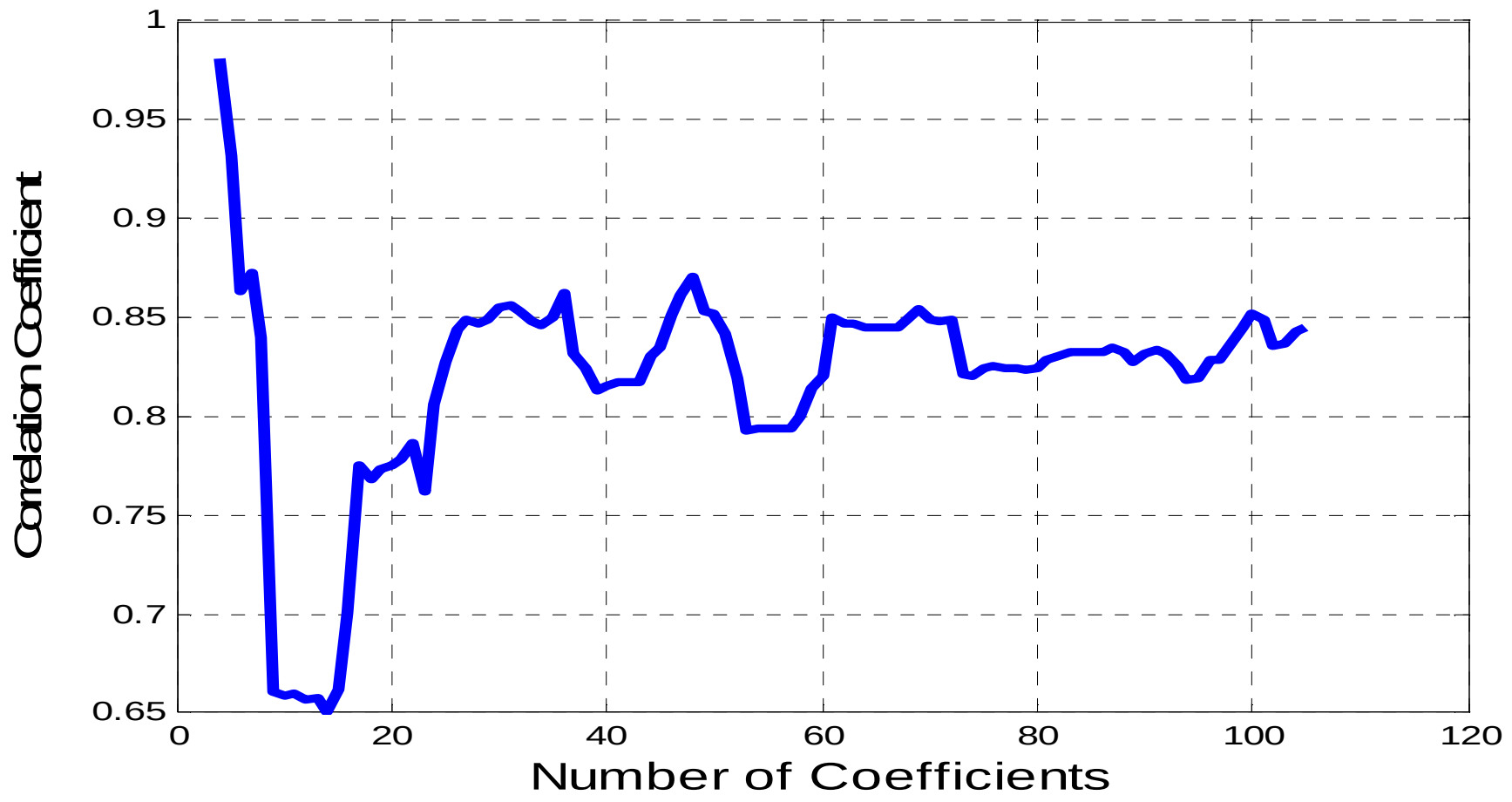

Figure 8. Correlation profile in terms of the number of coefficients of the adaptive filter. This helps in the design of a suitable predictor for the Brazilian Stock Market (PETR3) by adaptive filtering.

The initial region exhibiting very high correlation is not useful because it can only be achieved with a very small prediction window. The first-rate choice of the number of coefficients should be between 60 and 100 in this particular scenario.

Now, to evaluate the influence of length of the prediction window in the quality of the predictor, the front profile of the three-dimensional surface shown in Fig. 6 was drawn in Fig. 9. It shows the behavior of the maximum correlation in terms of window size. Clearly, the choice of a very short window leads to a high correlation (indeed, share quotations between successive days show a high correlation). But this is not of great significance in an investment strategy.

Surprisingly it appears that a choice of length of the window prediction around 10 days is quite inefficient. In contrast, prediction with a window of about 15-working days seems to be quite appropriate. Coincidentally, the length selected in the application shown in Section 3 was appropriate.

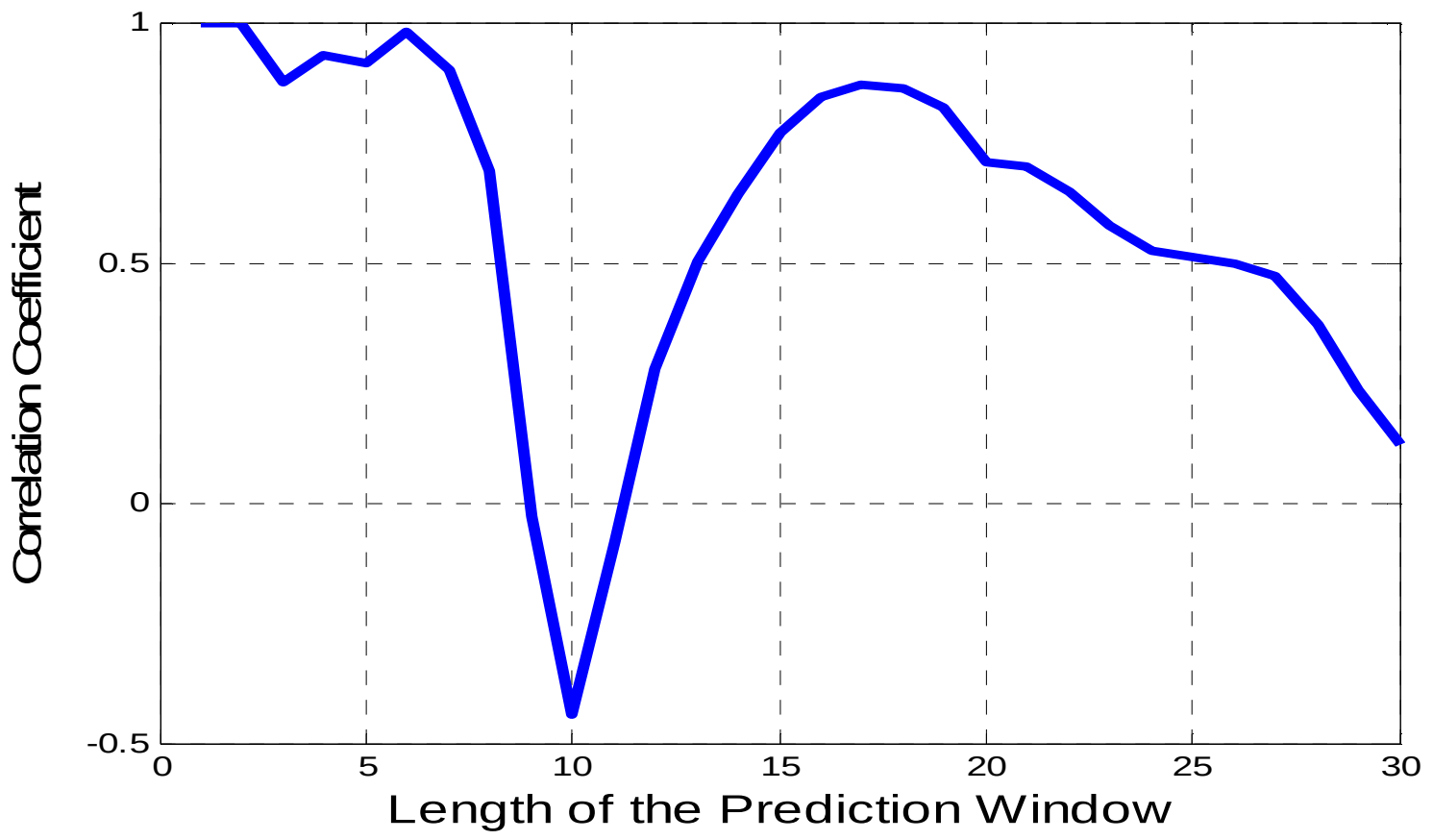

Figure 9. Correlation profile in terms of the size of the prediction window (from 1 to 30 opening days of the Brazilian stock market). This helps in the design of a suitable predictor for the PETR3 by adaptive filtering.

TABLE I. FILTER DESIGN FOR BRAZILIAN STOCK MARKET: INVESTIGATION OF THE INFLUENCE OF THE NUMBER OF COEFFICIENTS AND THE LENGTH OF THE WINDOW ON THE EARNINGS SUPPLIED BY THE PREDICTOR.

| *N* | *L* | *PD* | *SD* | *PV* | *SV* | *profit* |
|---|---|---|---|---|---|---|
| 60 | 20 | 2477 | 2482 | 37.77 | 39.75 | 5.24% |
| 65 | 19 | 2476 | 2482 | 37.86 | 39.75 | 4.99% |
| 70 | 18 | 2475 | 2486 | 37.33 | 41.05 | 9.97% |
| 75 | 17 | 2477 | 2485 | 37.77 | 41.00 | 8.55% |
| 80 | 16 | 2476 | 2486 | 37.86 | 41.05 | 8.43% |
| 85 | 15 | 2476 | 2485 | 37.86 | 41.00 | 8.29% |
| 90 | 16 | 2476 | 2486 | 37.86 | 41.05 | 8.43% |
| 95 | 17 | 2477 | 2487 | 37.77 | 41.09 | 8.79% |
| 100 | 18 | 2476 | 2483 | 37.86 | 39.86 | 5.28% |

Additionally, note also that a too long prediction window results in a lower correlation, as was expected. So, in this case, a margin of choice for the moment of prediction exists (typically between 15 and 20 days of opening) in order to achieve better results for the predictor.

Regarding to the sensibility of the prediction outcome, a more meticulous examination was then carried out and its results are listed in Table I. By varying the number *N* of FIR filter coefficients and the length *L* of the window prediction, we derived the profits shown in the last column of the table. The dates for the purchase and sale of the shares suggested by the algorithm (PD and SD, respectively) are also included. Assuming that an investor uses this strategy, its actual profit could be calculated by means of the historical series of quotations regarding this asset. The purchase and sale actual values of assets are indicated (PV and SV, respectively) as well as the corresponding actual profit that would be achieved in the "real world". Indeed, no one knows in advance the value of stock prices of assets in the days suggested by the action strategy of the algorithm (for buying and selling assets). However, the average earnings are always around 7% in the selected range of filter parameters, corroborating the robustness of the adaptive predictor offered in this paper. Recommended values for the length of the filter and the prediction window would be for example, in the vicinity of 80 and 16, respectively.

Both the length of the filter as the suggested window length was acceptable. Actually, FIR filters with less than 100 coefficients are not prohibitive complexity and are perfectly feasible. Moreover, a prediction horizon of less than a month is quite suitable for applications in the stock market.

## 5. Concluding Remarks

The adaptive filtering has attracted much attention, challenges and variety of possible applications. This paper investigates the application of adaptive filters in stock market prediction. Adaptive filtering was used to generate a forecast on the price quotation of Petrobrás shares traded on the Brazilian stock exchange, which is the main driver of Brazilian stock market. The actual valuation of the PETR3 share (Petrobrás ON) traded in the Brazilian Stock Market through adaptive filtering was investigated. It was shown that this approach furnishes on the average very substantial profit on the invested amount. The performance of the adaptive filter estimation was assessed, and a correlation measure was used to investigate the goodness of fit between the predicted signal and the actual market quotation. The influence of both the window length and number of coefficients of the FIR filter in the quality of the predictor were examined. It is fairly clear that the estimated profit (8.43%) on this investment strategy does not stand for a general behavior of the method, since it is merely a punctual estimation. Nevertheless, it suggests that the technique can provide high profits and also furnishes an order of magnitude of profits. A random selection of a large number of periods of investment (while buying and selling of shares) should be made. A deeper statistical study is being conducted to shed light on the potential of the technique. Therefore, this paper offers new predictors based on adaptive filtering as a decision making engineering tool on investment strategies to be adopted in the stock market. The results are very promising, requiring, nevertheless, a more careful evaluation. Nevertheless they unveiled a new set of applications of predictive filtering in the field of Economics.